\documentstyle[12pt,epsf]{article}

\def\abstracts#1{{
\centering{\begin{minipage}{16.2truecm}\vspace*{.1cm}
        \baselineskip=24pt
        \noindent\parindent=0pt #1
        \end{minipage}}\par}}

\newcommand{\lsim}{\buildrel < \over {_\sim}}
\newcommand{\ie}{{\it i.e.}}
\newcommand{\etal}{{\it et al.}}
\newcommand{\order}[1]{${\cal O}(#1)$}
\newcommand {\pom} {{I\hspace{-0.2em}P}}
\newcommand {\poml} {{I\hspace{-0.3em}P}}
\newcommand {\xpom} {\mbox{$x_{_{\pom}}$}}
\newcommand{\eq}[1]{Eq. (\ref{#1})}
\renewcommand{\thefootnote}{\fnsymbol{footnote}}
\begin{document} \begin{titlepage}

\setcounter{page}{1}
\thispagestyle{empty}
\rightline{\vbox{\halign{&#\hfil\cr
                 &NORDITA - 95/53 P\cr
                 &hep-ph/9507367\cr}}}
\vspace{0.2in}
\begin{center}

\large{\bf THE INTERPRETATION OF RAPIDITY GAPS AT HERA}\\
\bigskip
\vspace{0.3cm}
{\bf Paul Hoyer and C. S. Lam\footnote{Permanent address: McGill University,
Montreal, Canada}}\\
\vspace{.04in}
\it NORDITA, Copenhagen, Denmark\\
\vspace{0.4cm}
\end{center}
\begin{center}
ABSTRACT
\end{center}
\abstracts{
In leading twist deep inelastic $ep$ scattering, the virtual photon
interaction is fast compared to the time scale of soft color
rearrangement. We compare the Pomeron exchange model, in which a neutral
cluster is preformed, with a gluon exchange model, in which color is exchanged
after the hard interaction. We find several features of the DIS data and of
data on exclusive hard processes that favor a gluon exchange scenario. If
correct, the postulate of soft color interactions between the produced
($q\bar q$) system and the target has important implications for other
processes. In particular, this may explain the puzzles of charmonium
hadroproduction. }
\vskip0.3in

\renewcommand{\thefootnote}{\arabic{footnote}}
\end{titlepage}

\section{Introduction}

The discovery at HERA \cite{hera,hone,zeus} of Deep Inelastic Scattering
(DIS) events with a large rapidity gap between the particles produced in the
target (proton) and current (virtual photon) fragmentation regions raises
interesting questions concerning the principles of color neutralization in
hard scattering. The data rather convincingly demonstrate that the gap
events are of leading twist, \ie, they are not power suppressed  at large
photon virtualities $Q^2$.  Hence the virtual photon scatters off single
quarks and gluons in the target proton.

In leading twist scattering, the transverse momenta of the final state
partons are of \order{$Q$}. Since they eventually combine to color singlet
hadrons, there must be color exchange between the produced partons and the
proton remnant. This non-perturbative process is usually modelled in terms
of color strings. As two partons connected by a string fly apart, the
string breaks repeatedly and the rapidity interval between the partons is
populated with hadrons. Albeit heruristic, this picture has been tested
extensively and successfully, especially in $e^+e^-$ annihilations (where
no large rapidity gaps are observed).

In DIS the virtual photon takes a ``snapshot'' of the proton wave function
-- for $Q\gg \Lambda_{QCD}$ nonperturbative color exchange processes which
last 1 fm or longer are easily resolved. There are thus two principal
scenarios for the creation of rapidity gaps: The formation of color
neutral clusters can take place either {\em before} or {\em after} the hard
scattering. In the `Pomeron exchange' model \cite{IS,DL} the virtual
photon scatters off a preformed color neutral cluster (the Pomeron). In the
`Gluon exchange' models \cite{BH,EIR} the initial hard
scattering is quite similar in events with and without gaps. Following a
standard hard scattering $\gamma^*g \to q\bar q$, secondary soft gluon
exchange in the color field of the target is postulated to
transform the octet $q\bar q$ pair into a color singlet.

In this paper we want to discuss and compare these two quite different
approaches to gap dynamics. After a brief review of each model, we discuss
their consequences for several types of hard scattering, and compare with
available data. Naturally, neither of the models is likely to be fully
correct, but they represent the two main options for understanding the data.
Related models for the rapidity gaps in hard scattering processes
are given in \cite{diff}. For a recent review of hard diffraction and rapidity
gaps see Ref. \cite{VDD}.

\vspace{-1cm}
\begin{figure}[htbp]
\begin{center}
\leavevmode
{\epsfxsize=3.75truein \epsfbox{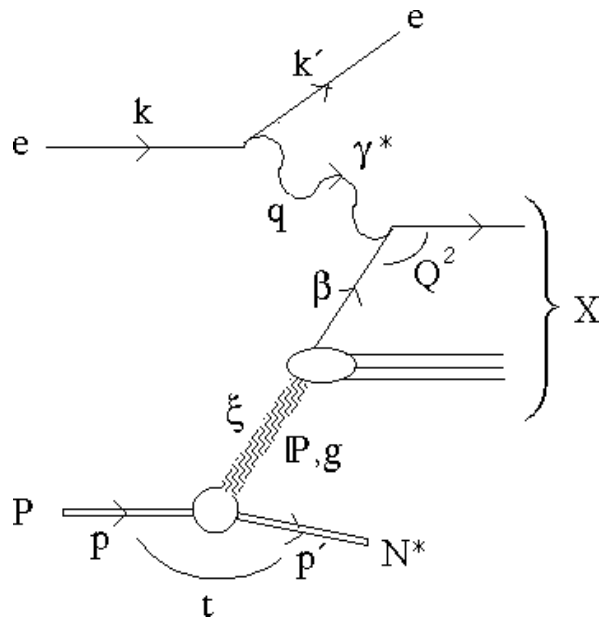}}
\end{center}
\caption[*]{Kinematics of the process $e(k)+p(p) \to e(k')+N^*(p')+X$. $q^2 =
(k-k')^2 = -Q^2$; $x_{Bj}=Q^2/2q\cdot p$; $t=(p-p')^2$; $\xi=q\cdot(p-p')
/q\cdot p$; $\beta = Q^2/2q\cdot(p-p')$.}
\label{fig1}
\end{figure}

\section{Two scenarios for DIS gaps}

The two types of models we shall consider are based on very different
dynamics, but their kinematics can be depicted using the same diagram
(Fig.~1) for the measured process $ep \to eN^*X$. In the present HERA data
the proton fragment $N^*$ (which may be a single nucleon) is not detected.
A rapidity gap is required between the proton beampipe (which contains
the $N^*$) and the hadrons comprising system $X$. The constraint that all
particles in $N^*$ escape detection limits the momentum transfer at the
nucleon vertex ($|t| \lsim 7$ GeV$^2$) \cite{hone} and mass of the proton
fragment ($M_{N^*} \lsim 4$ GeV) \cite{zeus}.

Depending on the model, the proton emits either a Pomeron $(\poml)$ or gluon
(g), which carries a small fraction $\xi$ of the proton momentum. The
photon then scatters on a parton carrying a fraction $\beta$ of the
Pomeron or gluon momentum. Assuming $|t| \ll Q^2, M_X^2$ both $\xi$ and
$\beta$ can be reconstructed from the measured quantities $x_{Bj}=\xi\beta$
and $M_X^2/Q^2=(1-\beta)/\beta$.

The two-step process of Fig.~1 implies that the measured ``diffractive''
structure function for events with rapidity gaps can be expressed as a
product
\begin{equation}
F_2^{gap}(x_{Bj},Q^2,\beta)=f_{i/p}(\xi,Q^2) F_2^i(\beta,Q^2) \label{fact}
\end{equation}
Here $f_{i/p}(\xi,Q^2)$ represents the probability for finding the emitted
object ($i=\poml$ or $g$) in the proton, while $F_2^i(\beta,Q^2)$ is the
structure function of this object. The dependence on $Q^2$ should be weak
(at most logarithmic) and an average has been taken over the (unmeasured)
virtuality $t$ of the Pomeron or gluon.

The HERA data \cite{hone,zeus} is consistent with the factorization
(\ref{fact}), and allows a separate measurement of the functions $f_{i/p}$
and $F_2^i$.

\subsection{Pomeron exchange}

The possibility of hard scattering on the Pomeron was proposed long before
the evidence for rapidity gaps at HERA \cite{IS,DL}. Soft diffractive
hadron-hadron scattering can be modelled by the exchange of a Regge
trajectory with vacuum quantum numbers. If this Pomeron
corresponds to an actual quark-gluon cluster in the hadron wave
function then it serves as a target for the virtual photon and will
give rise to DIS events with a rapidity gap.

Using Pomeron factorization, the momentum distribution $f_{\pom/p}$ in
\eq{fact} can be obtained from analyses of soft proton scattering. It
should be independent of $Q^2$ and (at $t=0$) its
dependence on $\xi = \xpom$ should be \cite{DLb,CHPWW}
\begin{equation}
f_{\pom/p}(\xpom,Q^2) \propto \xpom^{1-2\alpha_{\pom}(0)} \label{fpom}
\end{equation}
where $\alpha_{\pom}(0) \simeq 1.08$ is the intercept of the Pomeron Regge
trajectory. The data on soft scattering does not constrain the Pomeron
structure function $F_2^{\pom}(\beta,Q^2)$.
The basic assumption that the Pomeron is factorizable leads to quite powerful
predictions for a number of hard processes. We return to these below.

\subsection{Gluon exchange}

It is possible that the presence or absence of rapidity gaps in DIS is
determined by soft gluon dynamics long (\order{1\ \rm fm} in the target rest
frame) after the hard scattering \cite{BH,EIR}. Since the rapidity gaps are
observed at small values of $x_{Bj} \lsim 0.01$, a large fraction of the hard
scattering is due to $\gamma^*g \to q\bar q$. The function
$f_{i/p}(\xi,Q^2)$ in \eq{fact} should then be identified with the gluon
momentum distribution $g(x_g,Q^2)$ of fully inclusive DIS, with $\xi=x_g$.
The quark distribution in gluons $F_2^g(\beta,Q^2)$ is to leading order in
$\log Q^2$ given by the $g \to q\bar q$ splitting function \cite{BH}.

In this scenario the $q\bar q$ pair, which is produced as a color octet,
can turn into a singlet while passing through the color field of the
target. In this case no color string is stretched to the proton remnant, and
a rapidity gap is created. Assuming a statistical probability $P_1 \simeq
1/9$ for the $q\bar q$ to emerge as a singlet, Buchm\"uller and
Hebecker \cite{BH} found good agreement with the cross section and kinematic
distribution of rapidity gaps observed at HERA. The more detailed Monte
Carlo model of Ref. \cite{EIR} was also found to agree with the data.

It is important to note that the $q\bar q$ pair is formed at
transverse size $r_\perp \simeq 1/Q$, and expands with a velocity $v_\perp$ of
\order{Q/\nu} $ \ll 1$ in the proton
rest frame. Hence the pair remains compact while traversing the target and
as a color octet it interacts with soft target gluons. If such interactions
can indeed turn the quark pair into a color singlet then this has important
consequences for many other hard processes, including charmonium production.
We return to these below.

\section{Tests in hard diffraction}

\subsection{DIS gap events}

Many comparisons of the above (and related) models with data on DIS gap events
have recently been made \cite{hone,zeus,BH,EIR,diff,VDD,GS,zeusb}.
We note the following.

\subsubsection{$\xi$-dependence}
According to \eq{fpom}, Pomeron exchange predicts that the diffractive
structure function (\ref{fact}) is proportional to $\xpom^{-1.17}$ at small
$\xpom=\xi$ for all $Q$. For gluon exchange, the behavior should be given by
the gluon distribution, $\xi=x_g$ with
\begin{equation}
g(x_g) \propto x_g^{-a_g} \label{glue}
\end{equation}
The effective power $a_g$ found in
analyses of scaling violations in DIS \cite{honeb,zeusc,MSR} is in the
range $a_g=1.22 \ldots 1.35$ for $Q^2 = 4 \ldots 7$ GeV$^2$, and increases
with $Q^2$.

The $\xi$-dependence of the measured structure function (\ref{fact}) for
events with rapidity gaps has been parametrized as $f(\xi) \propto
\xi^{-a_{gap}}$. Averaged over $Q^2 > 8$ GeV$^2$ the result was
$a_{gap}=1.19 \pm .06 (stat) \pm .07 (syst)$ for the H1 data \cite{hone}
and $a_{gap}=1.30 \pm .08 (stat)~^{+~.08}_{-~.14}~(syst)$ for the ZEUS
data \cite{zeus}. In a direct determination of the gluon structure
function using (2+1) jet events, H1 obtained \cite{honec} for the gluon
structure function (\ref{glue}) an effective power $a_g=1.63 \pm .12$ at
$Q^2=30$ GeV$^2$. Interestingly, in this analysis the fraction $(8 \pm 2)$\%
of events with a rapidity gap was consistent with being independent of $x_g$ in
the measured range $.0019 < x_g < .18$.

Taken together, the $\xi$-dependence of the HERA gap events suggests a
somewhat steeper increase at small $\xi$ than expected for Pomeron
exchange, and is consistent with the gluon exchange model.

\subsubsection{$\beta$-dependence}
The measured structure function for DIS
gap events, \eq{fact}, is weakly dependent on $\beta$
\cite{hone,zeus}.This is in qualitative agreement with the gluon exchange
model. For Pomeron exchange further assumptions must be made to predict the
$\beta$-dependence.

\subsubsection{$t$-dependence}
For Pomeron exchange the $\poml pN^*$ vertex
should be the same as the one measured in soft hadron scattering. In
particular elastic recoil ($N^*=p$) should dominate inelastic. The
$t$-dependence for elastic recoil should be given by the proton form
factor \cite{DLb,CHPWW}.

In the gluon exchange model one expects the gluon virtuality to grow with
$Q^2$ as usual in DIS (in contrast to the Pomeron, whose virtuality is
independent of $Q^2$). However, the gluon momentum fraction $x_g$ is quite
small for the gap events. A gluon with $x_g = .01$ (close to the upper
experimental range) and virtuality 1 GeV$^2$ is separated from the proton by
4.6 units of rapidity. Hadronic fragments associated with gluons in this
rapidity range will typically be grouped with system $X$ in Fig. 1, rather
than with the
$N^*$. The removal of such a wee gluon from the proton (followed by color
compensation from soft gluon exchange) may imply a dominance of
elastic recoil ($N^*=p$) also in the gluon exchange model. Nevertheless,
the prediction is less precise than for Pomeron exchange.

\subsection{Hard exclusive diffraction}

Both the Pomeron and gluon exchange models, if correct for DIS gap events,
have implications for hard exclusive processes as well. This follows directly
from Pomeron factorization and, in the gluon case, from the fact that the
produced $q\bar q$ system remains compact while traversing the color field of
the target.

\subsubsection{$s$-dependence}
The $\gamma p \to J/\psi\,p$ cross section has been found \cite{zeusd} to
grow considerably with energy in the range 12 GeV $\le E_{CM}=\sqrt{s}
\le$ 114 GeV. Parametrized as $\sigma \propto s^\lambda$, the effective
exponent $\lambda \simeq .36 \pm .04$ is considerably larger than the $\lambda
\simeq .08$ expected for Pomeron exchange. A perturbative QCD calculation
\cite{ryskin,BFGMS} involving two hard gluon exchanges predicts the cross
section to increase like the square of the gluon structure function,
$[xg(x,m_{J/\psi}^2)]^2$ with $x\simeq m_{J/\psi}^2/s$. According to \eq{glue}
this would imply that $\lambda
\simeq 2(a_g-1) \simeq 0.44 \ldots 0.70$, which is rather larger than the
measured value, but consistent considering the uncertainties. The gluon
exchange model \cite{BH,EIR} for DIS gap events implies that only one gluon is
perturbative, hence $\lambda \simeq a_g-1$, closer to the data.

Recently, the $\gamma^* p \to \rho\,p$ cross section was measured in the HERA
energy range \cite{zeuse}. While the cross section increases only moderately
with energy at $Q^2=0$, as expected for Pomeron exchange, the increase is much
faster at $Q^2=8.8$ GeV$^2$ and $Q^2=16.9$ GeV$^2$. The conclusions are quite
analogous to those given above for $J/\psi$ production.

\subsubsection{$A$-dependence}
The nuclear target dependence of the cross section for incoherent $\rho$
meson electroproduction has been parametrized as
\begin{equation}
\sigma_{incoh}(\gamma^* A \to \rho A)\propto A^{\alpha(Q^2)} \label{cteq}
\end{equation}
The exponent has been found \cite{esixsixfive} to increase from $\alpha(0)
\simeq 2/3$ to $\alpha(5\ \rm GeV^2) \simeq 0.9$. This behavior is expected on
the basis of `color transparency' \cite{CT}, according to which a
transversally compact color singlet $q\bar q$ pair has a small reinteraction
probability in the nucleus.

It is a consequence of Pomeron factorization that all $A$-dependence of the
$\gamma^* A \to \rho A$ process must come from the $\poml AA$ vertex, which is
independent of $Q^2$. The observed $Q^2$-dependence of the power $\alpha$ in
\eq{cteq} thus breaks Pomeron factorization.

In the gluon exchange model \cite{BH,EIR} the $q\bar q$ pair is for high $Q^2$
created as a compact color octet which does interact repeatedly with
the color field of the target nucleus. Since little momentum is transferred
in these soft interactions, the nucleus is nevertheless effectively
transparent to the pair. Hence the effective power is $\alpha \simeq 1$ at
large $Q^2$, as observed. Note that for the nucleus $A$ to stay intact it is
important that the soft scattering can restore the color to the nucleon from
which it was removed by the initial perturbative gluon. This is possible
since the longitudinal momentum transfer in the rest frame of the nucleus is
very small, of \order{\Lambda_{QCD}^2/s}. The soft scattering is thus
longitudinally coherent over the whole nucleus.

The cross section for incoherent $J/\psi$ photoproduction has an
$A$-dependence corresponding to \eq{cteq} with $\alpha(m_{J/\psi}^2) \simeq
0.9$, \cite{esixnineone,nmc} which agrees well with the power obtained for
$\rho$ electroproduction. For the $J/\psi$ process the $A$-dependence is
known also for the coherent part of the cross section,
\begin{equation}
\sigma_{coh}(\gamma A \to J/\psi A)\propto A^{\alpha_{coh}} \label{acoh}
\end{equation}
with $\alpha_{coh}=1.40 \pm .06 \pm .04$ measured by E691 \cite{esixnineone}
and $\alpha_{coh}=1.19 \pm .02$ obtained by NMC \cite{nmc}.

For a factorizable Pomeron one would again expect $\alpha_{coh}\simeq 2/3$,
as observed in soft coherent scattering.

For a compact $q\bar q$ pair with a small (effective) reinteraction
probability in the nucleus, the forward ($t=0$) scattering is coherent over
the whole nucleus, and thus the forward cross section is proportional to $A^2$.
The requirement of coherence in the transverse direction implies a steepening
of the forward peak with increasing $A$, reducing the power to
$\alpha_{coh}=2-2/3=4/3$ for the full reaction cross section. This value is in
reasonable agreement with the power measured in the E691 and NMC experiments
\cite{esixnineone,nmc}.

\section{Hadroproduction of quarkonia}

The gluon exchange model \cite{BH,EIR} for DIS rapidity gaps postulates that
transversally compact $q\bar q$ pairs experience repeated soft color
interactions in the target. If correct, this has important consequences also
for the hadroproduction of heavy quarkonia. According to perturbative QCD (more
precisely, in the `color singlet model' \cite{BR}) quarkonia with charge
conjugation $C=-$ (such as the $J/\psi$) are produced in subprocesses like
$gg \to J/\psi g$. The extra gluon emission is required by the quantum
numbers of the $J/\psi$, and significantly reduces the cross section compared
to that for $C=+$ states like the $\chi_2$ $(J^{PC}=2^{++})$, which are
directly created through $gg \to \chi_2$.

As a matter of fact, the perturbative calculations seriously underestimate
the $J/\psi$ and $\psi'$ cross sections, whereas the prediction for $\chi_2$
is consistent with the measurements \cite{schuler,largept,VHBT}. Most
importantly, the discrepancies appear \cite{cdf} to be as large for the
bottomonium states ($\Upsilon$) as for the charmonia -- the new effect is
leading twist in the quark mass.

A $q\bar q$ pair which interacts repeatedly in the color field of the target
will not retain the quantum numbers of the initial perturbative gluons. Hence
there is no need for the emission of a perturbative gluon in the production
of $C=-$ states, and the cross sections of all charmonia are comparable. This
agrees with the trend of the data.

Note that a solution of the quarkonium hadroproduction puzzle along these
lines is quite different, in principle, from that of the `color octet'
model \cite{BF}. In the octet model, the gluon emitted in the final state is
related to the higher Fock states of the quarkonium. The emission happens
after a characteristic time in the rest frame of the quarkonium, hence
typically long after the heavy quarks have left the target. A minimal number
of such gluons are emitted, having a hardness related to the radius of the
charmonium state. The importance of this effect should decrease with the mass
of the heavy quark.

A further indicator of the dynamics of charmonium hadroproduction is provided
by the experimental observation that the $J/\psi$ \cite{biino} and $\psi'$
\cite{efivethreeseven} are produced unpolarized. In both the color singlet
\cite{VHBT} and color octet \cite{TV} models a transverse polarization is
predicted. A random color field may destroy the initial polarization of the
heavy quarks. An analogous effect of the vacuum color field on the
polarization of the annihilating quarks in the Drell Yan process has been
discussed in Ref. \cite{nachtmann}.

\section{Summary}

We have compared two alternative scenarios for the dynamics of rapidity gaps
in the final states of deep inelastic $ep$ collisions. The main
distinguishing characteristic of the two approaches is the time of formation
of the color singlet clusters (which are widely separated in rapidity). In the
Pomeron exchange model \cite{IS,DL} the virtual photon scatters off a
preformed neutral cluster. In the gluon exchange model \cite{BH,EIR} the hard
scattering is the same in events with and without gaps, and color is exchanged
afterwards. Both models have been shown to be in reasonable
agreement with the HERA data \cite{hera,hone,zeus}.

We discussed the models both in view of recent HERA data and in terms of
their predictions for hard exclusive diffractive processes. For DIS, a
distinguishing feature between the models is the behavior of the structure
function of the gap events at small values of the momentum fraction $\xi$
carried by the Pomeron or the gluon. There are indications
that the data favors or faster increase at small $\xi$ than expected for soft
Pomeron exchange, and is more consistent with the behavior of the gluon
structure function.

Analogously, the energy dependence of the exclusive process $\gamma p \to
J/\psi\, p$ is governed by the probability that the Pomeron or gluon carries a
small momentum fraction $\xi \simeq m_{J/\psi}^2/s$. The available
data again favors the faster increase at small $\xi$ given by gluon exchange.

The nuclear target $A$-dependence of hard exclusive processes is not
consistent with a factorizable Pomeron, but can be understood on the basis
of color transparency. Soft gluon interactions of compact $q\bar q$ pairs
in the color field of the target should not to upset the predictions
of color transparency.

The postulate \cite{BH,EIR} that compact $q\bar q$ pairs can have soft
interactions and change their color in the target will, if correct, have
important consequences also for other processes where the color quantum
numbers are essential. We discussed the case of quarkonium hadroproduction,
where severe discrepancies have been found between QCD calculations and the
data, which appear to be of leading twist in the quark mass. Soft color
interactions in the target could help explain why states such as the
$J/\psi$, $\psi'$ and $\chi_1$, which cannot be produced directly by the fusion
of two gluons, do not have suppressed production cross sections and are
produced unpolarized.

The HERA gap events have focused attention on our limited
understanding of how soft color interactions transform the perturbative parton
state into the observed hadron distributions. It will be interesting to study
experimentally which conditions can be imposed on the hadron distributions in
DIS without changing the $x_{Bj}$ and $Q^2$ dependence of the structure
function. The gap condition may be but one of many possibilities.

\vskip 1cm

{\bf \noindent Acknowledgements}

\noindent PH would like to thank W. Buchm\"uller and G. Ingelman for helpful
discussions.

\end{document}